\documentclass[%
superscriptaddress,
showkeys,
 preprint,
showpacs,preprintnumbers,
amsmath,amssymb,
aps,
prb,
]{revtex4-2}

\usepackage{graphicx}
\usepackage{dcolumn}
\usepackage{bm}
\usepackage{amsmath}
\usepackage{upgreek}
\usepackage{hyperref}

\begin{document}


\title{Broad-angle photon-pair generation from high-$Q$ flatband resonant 3R-MoS$_{2}$ metasurfaces}

\author{Tingting Liu}
\thanks{These authors contributed equally to this work.}
\affiliation{School of Information Engineering, Nanchang University, Nanchang 330031, China}
\affiliation{Institute for Advanced Study, Nanchang University, Nanchang 330031, China}

\author{Huifu Qiu}
\thanks{These authors contributed equally to this work.}
\affiliation{National Laboratory of Solid-State Microstructures and College of Engineering and Applied Sciences, Nanjing University, Nanjing 210093, China}

\author{Xintong Shi}
\affiliation{School of Information Engineering, Nanchang University, Nanchang 330031, China}

\author{Jumin Qiu}
\affiliation{School of Information Engineering, Nanchang University, Nanchang 330031, China}

\author{Shuyuan Xiao}
\email{syxiao@ncu.edu.cn}
\affiliation{School of Information Engineering, Nanchang University, Nanchang 330031, China}
\affiliation{Institute for Advanced Study, Nanchang University, Nanchang 330031, China}

\begin{abstract}
	
Spontaneous parametric down-conversion (SPDC) in ultrathin optical resonant metasurfaces offers a promising platform for integrated quantum light sources. However, conventional resonances suffer from steep momentum dispersion, limiting photon-pair generation to narrow excitation and emission angles. Here, we demonstrate a high-$Q$ flatband resonant 3R-MoS$_{2}$ metasurface for high-efficiency, broad-angle SPDC photon-pair generation. By combining Brillouin zone folding with in-plane symmetry breaking, we engineer a high-$Q$ flatband resonance in the near-infrared regime, maintaining a robust $Q$-factor across a wide range of incident angles. Exploiting the strong local electric field enhancement of resonance and the large second-order susceptibility of 3R-MoS$_{2}$, the metasurface boosts the degenerate SPDC pair-generation rate with orders of magnitude enhancement over an unpatterned film of identical thickness. Crucially, the near-zero band dispersion extends the effective emission angle to $\pm 5^{\circ}$ with negligible spectral drift $\sim 0.1$ nm and high directional tolerance. Under pulsed excitation with a finite spectral bandwidth, the metasurface maintains a substantial bandwidth-averaged photon-pair yield, while demonstrating strong spectral locking of the resonant emission channel against pump wavelength detuning despite the expected trade-off in generation efficiency. This work establishes a versatile strategy for momentum-independent quantum light sources, opening new avenues for scan-free quantum ghost imaging and quantum information processing.

\end{abstract}

\maketitle


\section{Introduction}

Photon pairs exhibiting quantum correlations are key resources for quantum information processing, optical quantum computing, and quantum ghost imaging \cite{Altmann2018, Moreau2019, elshaari2020hybrid}. Spontaneous parametric down-conversion (SPDC) serves as the primary route for generating such non-classical states of light. Conventionally, SPDC relies on bulk nonlinear crystals, but their strict phase-matching constraints and large footprints impede integration into compact on-chip quantum circuits\cite{Yesharim2025, fang2026mid}. Recently, van der Waals materials have emerged as promising low-dimensional platforms for integrated nonlinear photonic devices, offering high refractive indices, sub-wavelength thickness, and relaxed phase-matching constraints\cite{Zotev2025}. Among them, monolayer transition metal dichalcogenides (TMDs), possess exceptionally large second-order optical susceptibilities, but their atomic interaction thickness severely limits the photon-pair generation \cite{trovatello2021optical}. Furthermore, standard 2H-phase TMDs suffer from nonlinear signal cancellation in even-numbered layers due to restored inversion symmetry\cite{zhao2016atomically}. In contrast, 3R-stacked crystals break spatial inversion symmetry regardless of layer thickness. For example, the 3R-MoS$_{2}$ retains a non-vanishing $\chi^{(2)}$ tensor across hundreds of nanometers and also allows nonlinear response to scale constructively with the thickness of the stack \cite{xu2022towards}. Leveraging this, multilayer 3R-MoS$_{2}$ crystals has demonstrated drastically increased signal yield of nonlinear conversion processes including second-harmonic generation (SHG) and quantum photon-pair generation\cite{zograf2024combining, weissflog2024tunable}. Nevertheless, the conversion efficiency of a bare flake remains fundamentally constrained by the wavevector mismatch between the interacting fields. Although quasi-phase-matching can be pursued via complex 3R-MoS$_{2}$ stack engineering, it typically mandates micrometer-scale thicknesses that undermine ultra-compact integration\cite{trovatello2025quasi}. Consequently, realizing an efficient SPDC process over subwavelength thicknesses of 3R-MoS$_{2}$ is highly desirable for integration and compatibility with optical circuits. 

To address this challenge, optical metasurfaces incorporating arrays of nanoresonators have been actively explored to engineer the local density of optical states and amplify localized electromagnetic fields \cite{solntsev2021metasurfaces, Sharapova2023, li2024metasurface}. Because nonlinear optical conversion scales strongly with the local electric field intensity at both fundamental and nonlinear frequencies, resonant nanostructures can boost SPDC yields while preserving compact integration compared with unpatterned thin films. In this context, metasurfaces exploit high-$Q$ resonances, such as guided resonances\cite{zhang2022spatially, Qu2022, Ma2023}, Mie resonances\cite{Nikolaeva2021, santiagocruz2021photon, mazzanti2022enhanced, Prokhorov2026, Zhu2026}, and bound states in the continuum (BICs)\cite{parry2021enhanced, santiagocruz2022resonant, Noh2024}, to achieve strong electromagnetic field localization, and demonstrate significantly enhanced nonlinear optical conversion efficiency. In particular, the quasi-BIC resonances exhibiting exceptionally high $Q$-factors and intense local field enhancements\cite{koshelev2018asymmetric, huang2023resonant, Wang2024}, have proven to be powerful mechanism for various nonlinear processes\cite{liu2019high, xu2019dynamic, carletti2019high, Liu2025, liu2026high}. This concept has also been extended to nonlocal metasurfaces made of 3R-stacked MoS$_{2}$ flakes. Recent breakthroughs have showcased 3R-MoS$_{2}$ metasurfaces supporting quasi-BIC resonances yielding orders-of-magnitude enhancement in SHG and SPDC efficiency within subwavelength thickness compared with the unpatterned flake\cite{peng20253r, fan2025enhanced, Wang2025}. However, conventional quasi-BIC modes anchored at the $\Gamma$-point intrinsically exhibit pronounced dispersion, setting stringent limitations on the accessible photon momentum and the angular excitation bandwidth. On one hand, tight-focusing pump beams, which inherently possess a broad spatial angular spectrum, cannot efficiently couple into the narrow-band, angle-sensitive resonance\cite{Reva2026}. On the other hand, the generated photon pairs are tightly confined to a narrow emission angle, leading to severe angular selectivity and necessitating complex optical raster-scanning to perform wide-field quantum applications such as quantum ghost imaging\cite{Ma2025a, Ren2026}. 

In this work, we propose and demonstrate a high-$Q$ flatband resonant 3R-MoS$_{2}$ metasurface for high-efficiency and broad-angle photon pair generation. By introducing a lattice-constant doubling to fold the Brillouin zone and breaking the supercell symmetry, we fold the flat guided mode into the radiative continuum, harnessing the quasi-BIC transition at the $\Gamma$ point to open an accessible yet high-$Q$ radiation channel across the continuous flatband. Utilizing quantum-classical correspondence, our numerical calculations confirm that the high-$Q$ flatband resonance yields SPDC photon generation at the resonant wavelength with high spectral brightness across a remarkably wide angular emission range. The high-$Q$ flatband property enables strong field confinement under wide-angle illumination, extending the SPDC emission angular spectrum up to $\pm 5^{\circ}$ and enhancing the pair generation rate by four orders of magnitude larger than that of an unpatterned 3R-MoS$_{2}$ film. Our work establishes a versatile strategy for broad angular emission of photon pairs from van der Waals nanostructures and unlocking new possibilities for scan-free, wide-field quantum ghost imaging.

\section{High-$Q$ flatband resonance in 3R-MoS$_{2}$ metasurface}

Figure 1(a) schematically illustrates the broad-angle, resonantly enhanced SPDC photon-pair generation from a flatband 3R-MoS$_{2}$ metasurface. In the SPDC process, a high-energy pump photon $\omega_{\text{p}}$ spontaneously decays within the nonlinear medium into a pair of momentum-correlated, lower-energy signal $\omega_{\text{s}}$ and idler $\omega_{\text{i}}$ photons under energy and momentum conservation. To boost the pair generation probability over sub-wavelength thicknesses, the metasurface is engineered to increase the local density of optical states within the nonlinear medium at the generated photon frequency. As shown in the inset, the metasurface unit cell consists of a pair of asymmetric 3R-MoS$_{2}$ elliptical nanodisks supported by a SiO$_{2}$ substrate, with rectangular lattice constants $P_{x}=430$ nm. The elliptical nanodisks are with identical height $H=300$ nm and separated by $P_{x}-\Delta P=410$ nm, whereas the long and short axes of the two nanodisks with $r=165$ nm and $r-\Delta r=155$ nm are orthogonalized. The designed metasurface supports a high-$Q$ flatband resonance under transverse electric (TE) illumination, providing highly customizable resonant wavelengths and linewidths through parameter tuning. When generated photon wavelength matches the resonance condition determined by the metasurface, the photon-pair emission is resonantly enhanced. 

\begin{figure*}[htbp]
	\centering
	\includegraphics
	[scale=0.6]{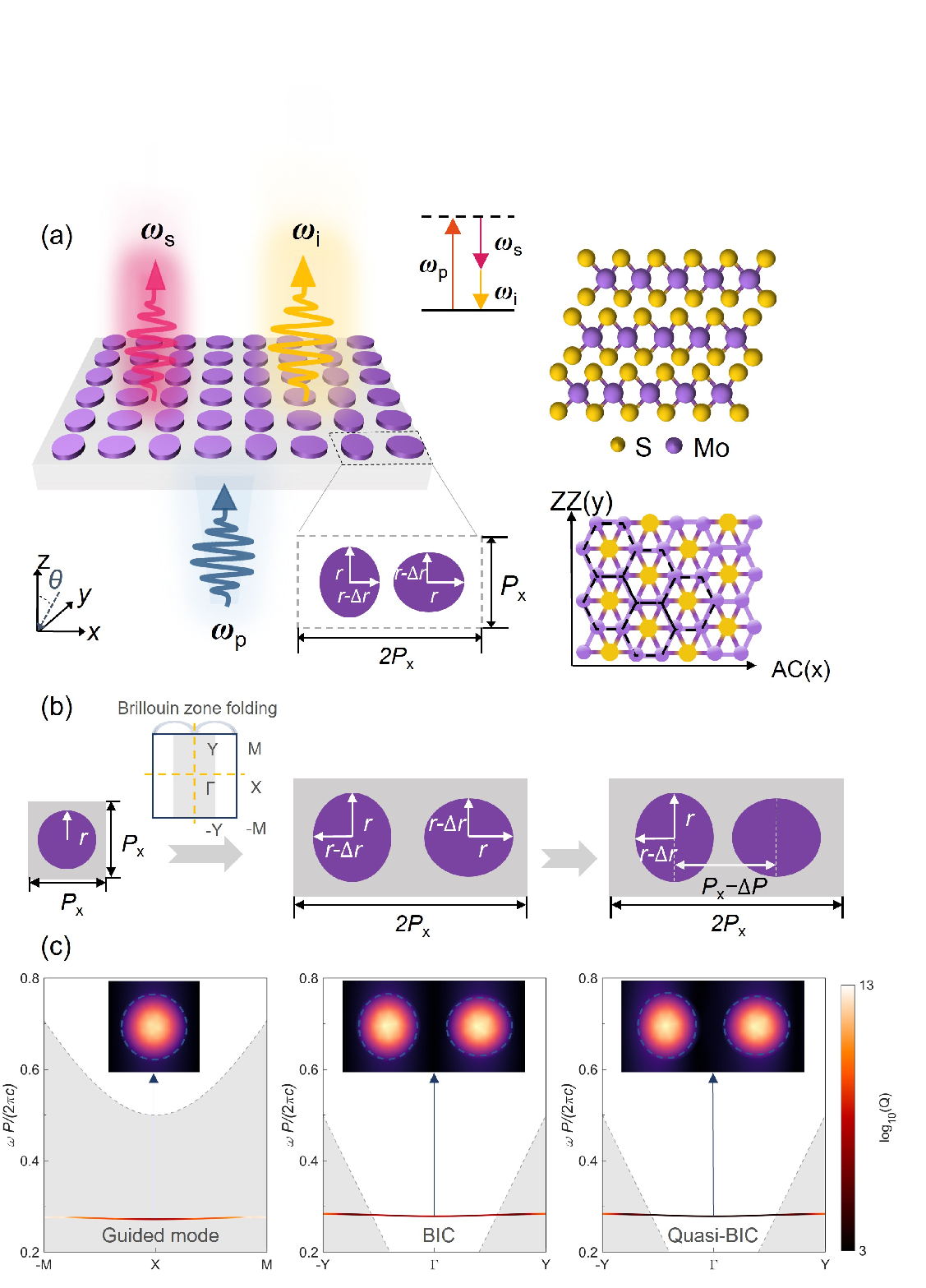}
	\caption{\label{Fig1} Concept and dispersion engineering of the high-$Q$ flatband resonance in a 3R-MoS$_{2}$ metasurface. (a) Schematic illustration of broad-angle, resonantly enhanced SPDC process for photon-pair generation from the flatband 3R-MoS$_{2}$ metasurface. Top inset: Energy-level diagram of SPDC where high-energy pump photons ($\omega_{\text{p}}$) down-convert into signal ($\omega_{\text{s}}$) and idler ($\omega_{\text{i}}$) pairs. Right inset: atomic crystal structure of non-centrosymmetric 3R-MoS$_{2}$. Bottom inset: Metasurface unit cell with orthogonal asymmetric 3R-MoS$_{2}$ elliptical nanodisks supported on a SiO$_{2}$ substrate. (b–c) Formation mechanism of the flatband resonance. (b) Brillouin zone folding achieved by doubling the lattice period along the $x$-direction ($P_{x}$), and in-plane asymmetry perturbation breaking $C_{2v}$ symmetry, transforming the symmetry-protected BIC at the $\Gamma$ point into a leaky quasi-BIC, thereby forming a continuous high-$Q$ flatband resonance. (c) Photonic band structure during the transformation process, and corresponding magnetic field amplitude $|H_{z}|$ within the $x$-$y$ plane of the structure.}
\end{figure*}

To achieve the flatband regime by engineering modal dispersion, we implement a supercell symmetry-breaking strategy via Brillouin zone folding. For a metasurface with a single circular nanodisk per unit cell with a period of $P_{x}$, the structure supports a guided mode at the $X$ point of the first Brillouin zone, as shown in Figs. 1(b) and 1(c). This mode exhibits a nearly flat dispersion around the $X$ point along the -M$-$X$-$M direction, with only weak frequency variation over a broad range of in-plane wavevectors. Following modulating the geometries of adjacent nanodisks to break the original translational symmetry, we introduce a supercell with a doubled lattice period of $P_{x}$ along the $x$-direction. Accordingly, the first Brillouin zone is halved, and the flat guided mode originally situated below the light cone around the $X$ point is folded back into the radiation continuum at the $\Gamma$ point, forming a genuine symmetry-protected BIC at $k_\parallel = 0$. Through this Brillouin-zone-folding process, the intrinsic flat dispersion around the $X$ point is mapped to the vicinity of the $\Gamma$ point, yielding a folded continuous guided-mode resonance band with a near-zero group velocity $\partial\omega/\partial k_\parallel \approx 0$. Because a true symmetry-protected BIC remains completely decoupled from normal-incident radiation, we further introduce an in-plane structural asymmetry via the gap perturbation $\Delta P$ between adjacent nanodisks. This perturbation breaks the spatial symmetry, transforming the isolated BIC at the $\Gamma$ point into an accessible leaky quasi-BIC and opening an efficient radiation channel to free space. Consequently, this mechanism enables free-space excitation of the folded mode while maintaining strong electromagnetic confinement, establishing a continuous high-$Q$ flatband resonance across a broad range of in-plane wavevectors. This combination of accessible radiative coupling and ultra-flat dispersion lays a solid foundation for broad-angle, angle-insensitive photon-pair generation.

It should be noted that Brillouin zone folding itself merely relocates an existing momentum state to the $\Gamma$ point without fundamentally altering its original dispersion curvature. To achieve the near-zero energy dispersion, the parent guided mode must be strictly engineered prior to folding. By optimizing the lattice constant $P_x$ and the nanodisk radius $r$, we effectively minimize the inter-cell electromagnetic coupling as detailed in Section 2 of the Supplemental Material, which inherently flattens the parent band along the folding direction. Subsequently, Brillouin zone folding maps this meticulously engineered low-dispersion band to the $\Gamma$ point, where targeted symmetry breaking transforms the bound state into a high-$Q$ flatband resonance.

\section{Broad-angle photon-pair generation}

Figure 2(a) displays the simulated angle-resolved transmission spectra under $y$-polarized plane wave excitation in the $y$-$z$ plane. The sharp resonance branch around 1542.4 nm corresponds to the engineered high-$Q$ flatband mode under TE polarization, whereas the resonance around 1423.3 nm represents a conventional highly dispersive mode. Remarkably, as the angle of incidence varies up to 15$^{\circ}$, the resonant wavelength of the flatband mode remains almost invariant around 1542 nm, contrasting starkly with the rapid spectral shift observed in the dispersive branch. Eigenmode analysis for different asymmetry perturbations in Section 3 of Supplemental Material validates the broad-angle high-$Q$ nature of the flatband resonance. We also apply the standard Fano-fitting procedure to the flatband transmission spectra in Section 5 of Supplemental Material. It reveals that the resonance maintains a robust $Q$-factor of $\sim 3\times10^{3}$ over the entire angular range from 0$^{\circ}$ to 15$^{\circ}$. This combination of ultra-flat dispersion and persistent high $Q$-factors confirms that the metasurface provides uniform and strong electromagnetic field localization across a broad momentum spectrum.

To predict photon-pair emission of SPDC process, we numerically calculate its classical reverse process of sum-frequency generation (SFG) within the metasurface. For the 3R-MoS$_{2}$ nanoresonators, the symmetry axis along the $x$-direction is aligned with the crystallographic armchair (AC) direction of the MoS$_{2}$ crystal, whereas the $y$-axis aligns with the zigzag (ZZ) direction, as shown in the inset of Fig. 1(a). Since the 3R-MoS$_{2}$ is non-centrosymmetric and possesses crystal lattice with rotational symmetry around the $z$-axis, its $\chi^{(2)}$ tensor exhibits non-zero in-plane elements \cite{munkhbat2022optical, weissflog2024tunable}. The non-zero tensor elements are given by
\begin{equation}
	\chi_{\alpha\beta\gamma}^{(2)}=\chi_{yyy}^{(2)} = -\chi_{yxx}^{(2)} = -\chi_{xxy}^{(2)} = -\chi_{xyx}^{(2)}.
\end{equation}\label{eq1}
This tensor couples the electric field components at down-converted frequencies, i.e., $\omega_{\text{s}}$, $\omega_{\text{i}}$, with polarization indices $\alpha$, $\beta$, to a higher-frequency pump field with $\omega_{\text{p}}=\omega_{\text{s}}+\omega_{\text{i}}$ and polarization index $\gamma$. According to the Lorentz reciprocity theorem and quantum-classical correspondence, this nonlinear polarization mechanism governs classical frequency-mixing processes such as SHG and SFG, and directly dictates the probability amplitude of spontaneous photon-pair generation in reverse process of SPDC. The full-wave linear and nonlinear numerical simulations are performed in the undepleted pump approximation\cite{feng2023enhanced}. We first compute the localized linear fundamental electric fields within the 3R-MoS$_{2}$ nanostructures, from which the induced nonlinear polarization for SFG process are extracted to calculate the far-field SFG intensity. Optical constants of 3R-MoS$_{2}$ and detailed numerical methods are provided in Sections 1 and 7 of Supplemental Material. 

To quantify the resonant enhancement of the nonlinear response, we evaluate the SFG conversion efficiency, defined as $\eta=P_{\text{SFG}}/(P_1+P_2)$, under dual $y$-polarized fundamental excitations in Fig. 2(b). Here, the two degenerate fundamental beams emulate the signal and idler photons involved in the time-reversed SPDC process, sharing identical spatial and polarization properties. Such degenerate SFG process can be considered as a SHG process. As the fundamental excitation wavelength approaches 1542.4 nm, a dramatic enhancement in SHG efficiency is observed, achieving nearly three orders of magnitude increase compared to off-resonance wavelengths and an unprecedented four orders of magnitude boost over an unpatterned 3R-MoS$_{2}$ film of identical thickness. This massive nonlinear enhancement originates directly from the intense local field confinement provided by the high-$Q$ quasi-BIC mode. At resonance, the spatial field distribution in Section 6 of Supplemental Material reveals a maximum local intensity enhancement $|E/E_{0}|^{2}$ of up to $3.2\times 10^{3}$ for the $y$-polarized field component inside the 3R-MoS$_{2}$ nanostructures. Therefore, a high local intensity enhancement greatly improves the classical degenerate SFG process. Because the SPDC generation rate scales with local field intensity enhancement at the signal and idler frequencies, this huge field localization could also significantly boost the pair-generation probability across the resonant regime.

\begin{figure}[htbp]
	\centering
	\includegraphics
	[scale=0.6]{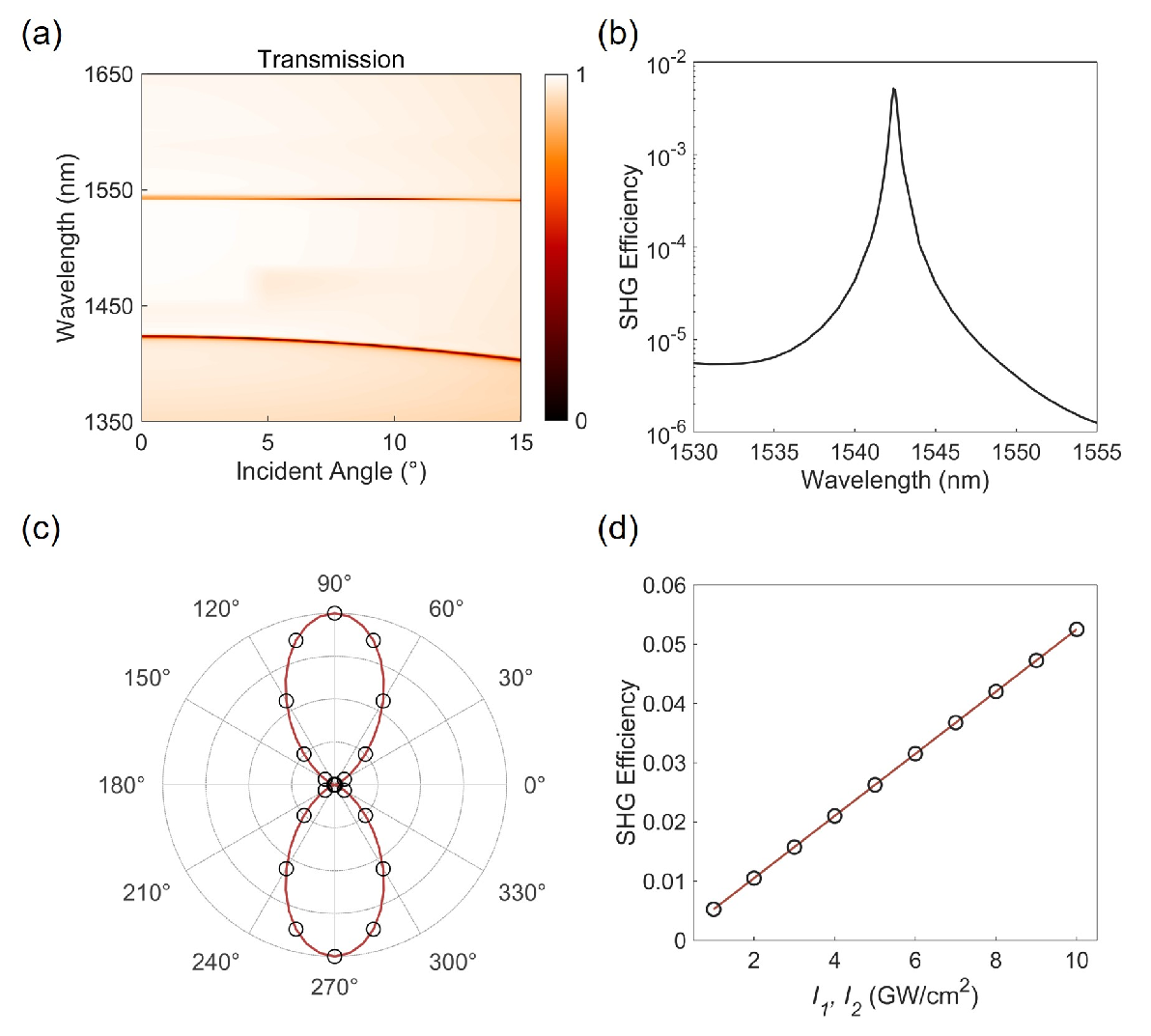}
	\caption{\label{Fig2} Transmission spectra and resonantly enhanced classical SHG. (a) Angle-resolved transmission spectra under $y$-polarized TE excitation across incident angles from $0^{\circ}$ to $15^{\circ}$. (b) SHG conversion efficiency $\eta=P_{\text{SFG}}/(P_1+P_2)$ as a function of fundamental wavelength under degenerate $y$-polarized excitation. (c) Dependence of SHG conversion efficiency on the excitation polarization angle. (d) SHG output efficiency as a function of fundamental excitation intensity $I_{1}$, $I_{2}$ at the resonance wavelength.}
\end{figure}

It is important to note that, introducing a high-$Q$ resonance with a preferred field orientation substantially modifies the polarization state of the generated photons. In unpatterned 3R-MoS$_{2}$ thin films, the nonlinear tensor symmetry directly supports the generation of polarization-entangled photon pairs. For example, the crystal symmetry of a multilayer 3R-stack of MoS$_2$ enables the generation of polarization-entangled Bell states without additional components and provides tunability by simple control of the pump polarization\cite{weissflog2024tunable}. But the case is different in the resonant system\cite{Weissflog2024, Jia2025, Ma2025, fan2025enhanced}. In our metasurface platform, the localized high-$Q$ flatband mode selectively amplifies the local density of optical states along the $y$-axis. Consequently, the probability of generating $y$-polarized photon pairs is selectively heightened relative to $x$-polarized channels, leading to nonentangled $|VV\rangle$ state. This resonant polarization selectivity is further confirmed by the pump-polarization dependence of the SHG signal shown in Fig. 2(c). The SHG output reaches its maximum when the excitation polarization aligns perfectly with the $y$-axis, i.e., polarization angle $\theta=90^{\circ}$ in the figure, matching the dominant mode polarization of the flatband resonance. In Fig. 2(d), we present the SHG efficiency dependence on the pump intensity at the resonance wavelength. It is illustrated that the generated nonlinear signal exhibits a strict linear scaling behavior with respect to the fundamental excitation power density. This verifies that the frequency-conversion process operates strictly within the undepleted nonlinear regime.

To quantitatively evaluate the quantum photon-pair emission from the 3R-MoS$_{2}$ metasurface, we employ the quantum–classical correspondence principle, calculating the SPDC pair-generation rate via classical SFG efficiency as\cite{marino2019spontaneous, parry2021enhanced, jin2021efficient, liu2024efficient} 
\begin{equation}
	\frac{dN_{\text{pair}}}{dt} = 2\pi c \frac{\lambda_{\text{p}}^{4} \Delta \lambda}{\lambda_{\text{s}}^{5} \lambda_{\text{i}}^{3}} I_{\text{p}} \Xi_{\text{SFG}} .
\end{equation}\label{eq2}
The details for the calculation can be seen in the Section 7 of Supplemental Material. Fig. 3(a) illustrates the spectral density map of the biphoton generation rate across $(\lambda_s,\lambda_i)$ wavelength space, derived from the independent dual-wavelength sweep in the reverse classical SFG process. In this representation, the corresponding pump wavelength dynamically satisfies the energy conservation condition $\lambda_\text{p} = (\lambda_s^{-1} + \lambda_i^{-1})^{-1}$ at each point. When both generated photons simultaneously approach the high-$Q$ resonance, $\lambda_{\text{s}} = \lambda_{\text{i}}$ = 1542.4 nm, a prominent emission hotspot spanning a narrow bandwidth is observed. Furthermore, two orthogonal high-rate bands intersect at this central hotspot, corresponding to spectral channels where either the signal or idler photon independently couples into the high-$Q$ flatband mode. This multi-channel enhancement is rooted in the significantly increased local density of optical states provided by the nonlocal high-$Q$ resonance, which dramatically boosts the biphoton generation rate and yields ultrahigh spectral brightness.

We next investigate the nondegenerate SPDC regime by detuning the pump wavelength away from the exact half resonance condition $\lambda_{\text{p}} \neq \lambda_{\text{res}}/ 2$. Fig. 3(b) presents the nondegenerate SPDC generation rate across a broad range of pump and signal wavelengths. While the absolute maximum pair-generation rate is attained under the fully degenerate condition, shifting the pump wavelength breaks the spectral degeneracy, generating twin emission trajectories. Specifically, one photon remains spectral-pinned at the flatband resonance, whereas its companion photon is emitted at an energy-conserved wavelength governed strictly by: $\lambda_{\text{i}} = \left(\frac{1}{\lambda_{\text{p}}} - \frac{1}{\lambda_{\text{s}}}\right)^{-1}$. Because the SPDC process intrinsically creates time-correlated photons in pairs, the signal and idler channels exhibit identical generation rates, forming symmetric, dual-peak spectral features during the pump tuning process.

To explicitly reveal the resonant effect on the spectral properties of the generated photon pairs, Figs. 3(c) and 3(d) present the biphoton emission spectra for degenerate and nondegenerate cases, corresponding to the dashed cut-lines in Fig. 3(b). Under resonant degenerate pumping of wavelength $\lambda_{\text{p}}$=771.2 nm, both the signal and idler photons are emitted at the high-$Q$ resonant wavelength of 1542.4 nm. As shown in Fig. 3(c), the emission photons exhibit a producing a sharp emission peak reaching up to $3.86\times 10^3$ Hz, an enhancement of four orders of magnitude compared to an unpatterned 3R-MoS$_{2}$ flake. Notably, the biphoton spectrum exhibits a remarkably narrow bandwidth of $\sim0.47$ nm. In Fig. 3(d), we adopt the nonresonant pump excitation at wavelength of $\lambda_{\text{p}}=769.4$ nm. Conversely with the degenerate pumping, the emission spectrum splits into two distinct peaks, with one photon has a wavelength around the resonance, and the other photon is around 1535.2 nm. Due to the energy conservation and the nature of the SPDC process, the signal and idler photons are generated in pairs with equal probability. The identical emission rates and tight spectral correlations between these nondegenerate photon pairs offers significant advantages for compact quantum information processing platforms.

\begin{figure}[htbp]
	\centering
	\includegraphics
	[scale=0.6]{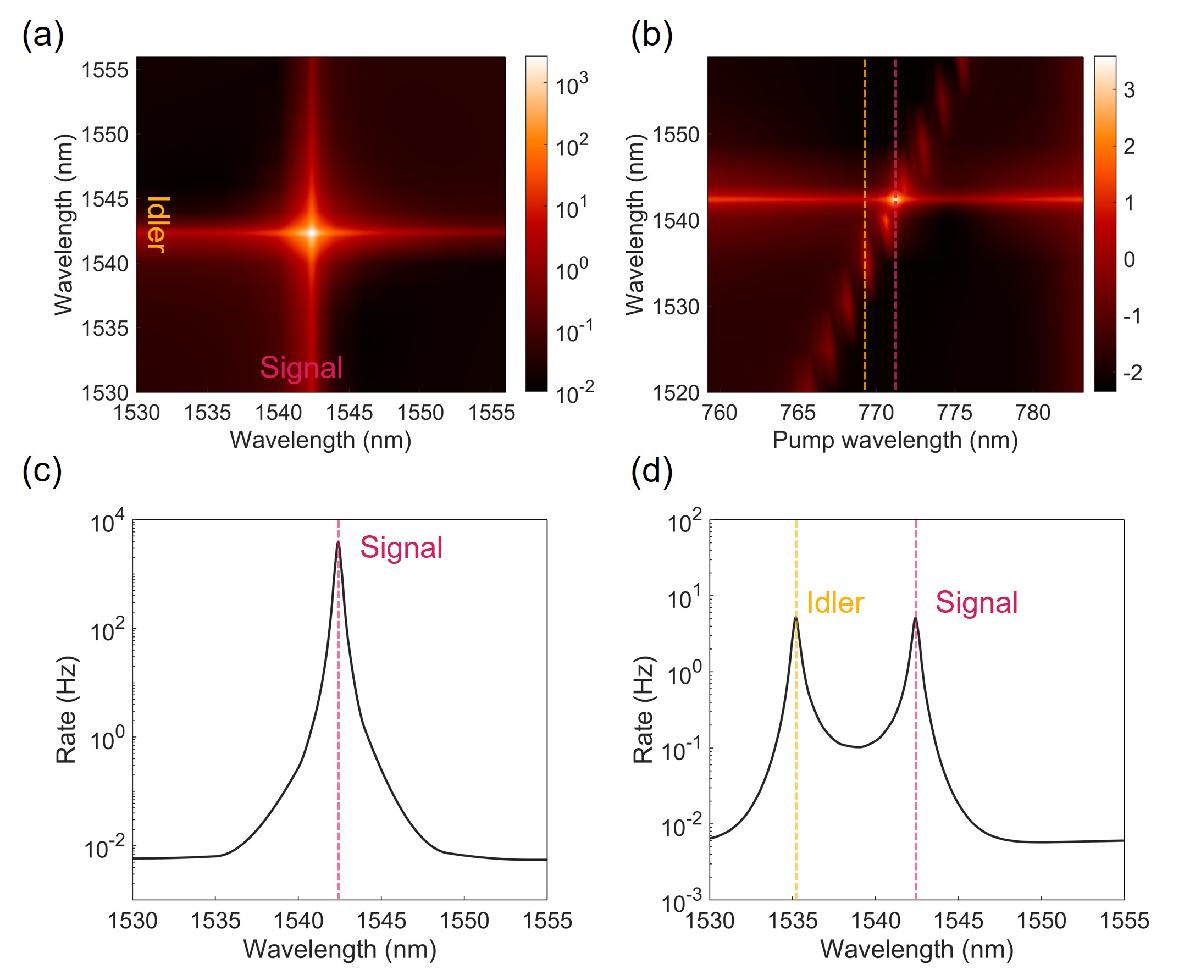}
	\caption{\label{Fig3} SPDC photon-pair generation rates and spectral characteristics. (a) Pair-generation rate map as a function of signal ($\lambda_{\text{s}}$) and idler ($\lambda_{\text{i}}$) wavelengths. A bright, resonantly enhanced hotspot appears at $\lambda_{\text{s}}=\lambda_{\text{i}}=1542.4$ nm. (b) Pair-generation rate map under varying pump wavelengths ($\lambda_{\text{p}}$). (c) Degenerate biphoton emission spectrum along the resonance line $\lambda_{\text{p}}=771.2$ nm exhibiting a sharp peak rate. (d) Nondegenerate biphoton emission spectrum under detuned pump excitation $\lambda_{\text{p}} = 770.6$ nm, demonstrating symmetric dual-peak splitting into flatband resonant and energy-conserved emission channels at $\lambda_{\text{s}}=1542.4$ nm and $\lambda_{\text{i}}=1535.2$ nm, respectively.}
\end{figure}

To elucidate the role of flatband resonance in enhancing quantum photon-pair rates and broadening the angular spectra, we analyze the dependence of the SPDC generation rate on the emission angle, as depicted in Fig. 4(a). Under degenerate pump excitation, $\lambda_{\text{p}} = \lambda_{\text{res}} / 2$, both signal and idler photons are emitted at the resonantly enhanced wavelength $\lambda_{\text{res}}$. Notably, the biphoton generation rates exhibit remarkable resonant enhancement across a wide angular spectrum, while maintaining an exceptionally stable central wavelength. Specifically, as the emission angle spans from $0^{\circ}$ to $5^{\circ}$, the spectral position of the maximum emission rate undergoes a negligible shift from 1542.4 nm to 1542.3 nm, which corresponds to less than $20\%$ of the resonance linewidth around 0.5 nm. Concurrently, the emission rate exhibits high angular tolerance, with the pair-generation rate at $5^{\circ}$ sustaining a high level ($\sim$ 1971 Hz) that is merely approximately half of the normal-incidence maximum at 1542.4 nm ($\sim$ 3867 Hz). Even when the emission angle extends to $10^{\circ}$, the spectral position of the maximum emission rate at 1541.9 nm exhibit only a shift of 0.5 nm relative the emission wavelenth at $0^{\circ}$. This angular resilience directly confirms that the flat energy dispersion effectively broadens the spatial acceptance cone for SPDC pair generation without sacrificing spectral purity. This robust performance across broad emission angles stems directly from the persistent high-$Q$ factor localized across the momentum landscape, as manifested by the Fano fitting of the angle-resolved transmission spectra in Section 5 of Supplemental Material.

\begin{figure}[htbp]
	\centering
	\includegraphics
	[scale=0.6]{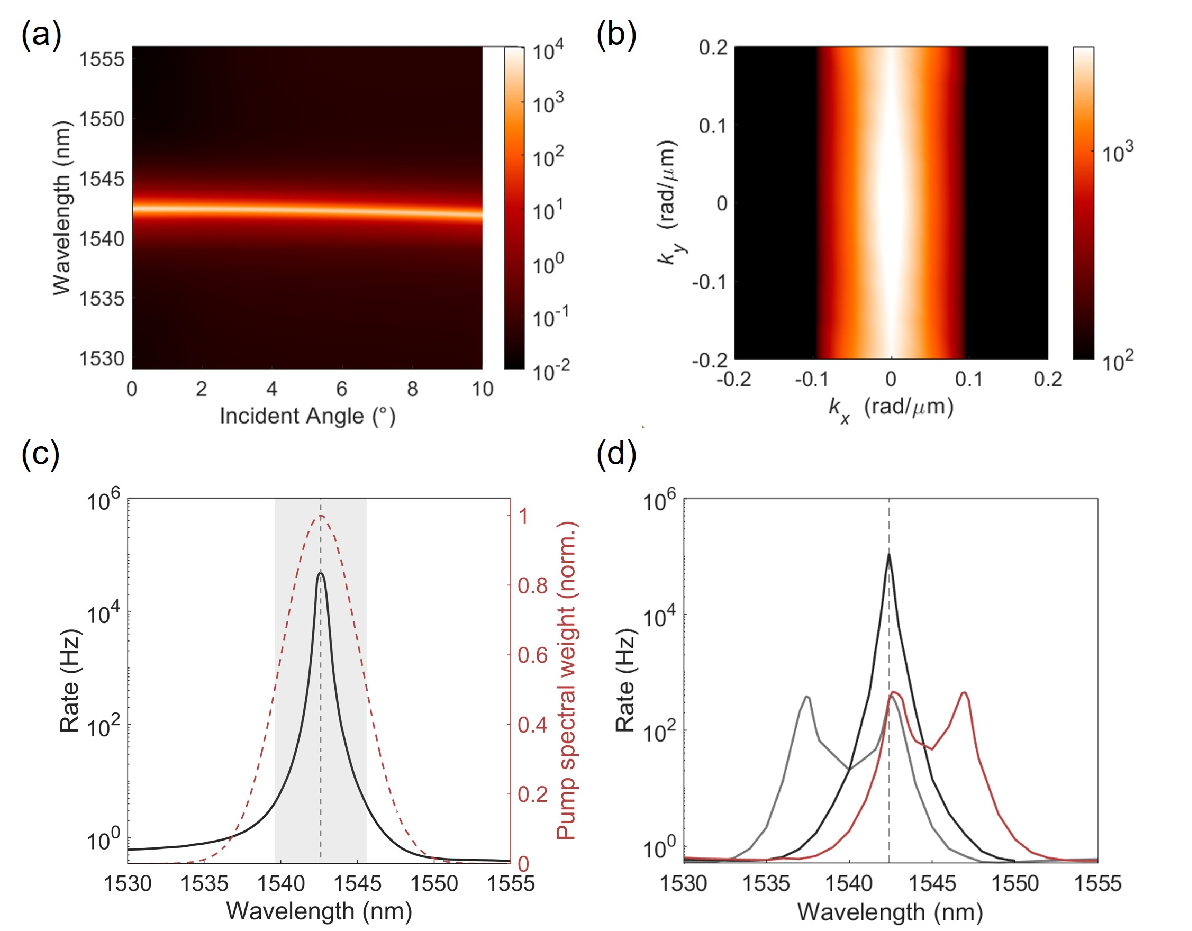}
	\caption{\label{Fig4} Broad-angle momentum independence, spatial radiation profile, and pulsed-pump robustness. (a) SPDC emission rate dependence on the emission angle $\theta$ up to $10^{\circ}$ under degenerate excitation. (b) Two-dimensional momentum-space radiation map of generated photon pairs of $|VV\rangle$ state at resonance, showing an extended 1D high-intensity plateau along the $k_y$ direction while rapidly decaying along the $k_x$ direction within $|k_{x,y}| \le 0.2$ rad/$\upmu$m. (c) Momentum-integrated SPDC rate as a function of $2\lambda_{\text{p}}$ (black solid curve) overlaid with the normalized Gaussian spectral profile of a 6-nm FWHM pulsed pump (red dashed curve). (d) Biphoton emission spectra under slight pump detunings. The primary peak remains strictly locked at 1542.4 nm for degenerate process under pump wavelength of 771.2 nm (black curve), and for nondegenerate cases under detuned pump wavelengths of 770 nm (gray curve) and 772.4 nm (red curve), while the energy-conserved companion photon channels manifest as symmetric spectral peaks.}
\end{figure}

We further characterize the spatial radiation properties of the generated quantum states. Fig. 4(b) displays the two-dimensional momentum-space ($k_{x}, k_{y}$) emission map calculated under $y$-polarized pump at resonance. In this regime, the generated signal and idler photons preserve the polarization of the pump beam, giving rise to a co-polarized photon-pair state $|VV\rangle$. Notably, the momentum-space emission exhibits a profound in-plane anisotropy, forming an extended 1D high-intensity plateau along the $k_y$ direction while rapidly decaying along the $k_x$ direction within $|k_{x,y}| \le 0.2$ rad/$\upmu$m. This highly anisotropic emission profile is the direct physical consequence of our unidirectional symmetry-breaking strategy. Because the lattice doubling and gap perturbations are applied along the $x$-axis, the resulting flatband maintains near-zero dispersion along $k_y$ but exhibits steep dispersion along $k_x$, as discussed in Section 4 of Supplemental Material. On this basis, we further investigate the usable photon-pair generation rate under tightly focused Gaussian-beam excitation in Section 8 of Supplemental Material. Compared with the conventional dispersive mode, the flatband resonance improves the practical usable photon-pair generation rate.

While ideal monochromatic excitation is advantageous for generating quantum states with high spectral purity, practical quantum optical systems often employ pulsed pump sources with finite spectral bandwidths\cite{Li2026}. To evaluate the practical pair-generation performance of the high-$Q$ flatband metasurface under realistic pulsed pumping, we simulate SPDC emission driven by a finite-bandwidth Gaussian pulsed pump at the exact same average pump power as the ideal monochromatic case. The detailed procedures are provided in Section 9 of the Supplemental Material. As shown in Fig. 4(c), we map a Gaussian pump spectrum with 3 nm full-width at half-maximum (FWHM) onto the degenerate signal wavelength axis, corresponding to a 6 nm FWHM envelope centered at $\lambda_\text{s}$= 1542.4 nm. Under ideal monochromatic excitation at exact resonance, integrating the generation rate over the central flatband momentum plateau $\vert{}k_{x,y}\vert{} \le 0.1\text{ rad}/\mu\text{m}$ yields a peak value of $4.75\times 10^4$ Hz. However, because the 6 nm pump bandwidth substantially exceeds the narrow resonance linewidth $\sim$ 0.47 nm, only the central spectral fraction of the pump pulse couples efficiently into the doubly resonant channel. Averaging across the off-resonant spectral wings inevitably reduces the effective bandwidth-weighted generation rate to $4.93\times 10^3$ Hz. Despite this fundamental physical trade-off between peak resonant enhancement and pump-bandwidth acceptance inherent to high-$Q$ resonators, the metasurface still preserves a substantially high usable pair yield owing to the strong local field enhancements provided by the flatband dispersion.
 
To investigate the stability of the generated photon pairs against pump laser fluctuations, we examine the signal emission spectra under slight pump detunings. Fig. 4(d) presents the signal emission spectra under slight pump detunings, with a $\pm 1.2$ nm shift relative to wavelength $\lambda_{\text{p}} = 771.2$ nm. Different from the broad dual-peak splitting observed under large wavelength separations as discussed in Fig. 3(d), we compare the emission spectra for different pump wavelengths over the central flatband momentum plateau $\vert{}k_{x,y}\vert{} \le 0.1\text{ rad}/\mu\text{m}$ as a function of the signal wavelength condition. As depicted in Fig. 4(d), the SPDC process under detuned excitation transitions from a doubly resonant degenerate regime into a singly resonant nondegenerate regime. Driven by the highly concentrated local density of optical states provided by the flatband resonance, one emitted photon consistently remains spectrally pinned at the resonance wavelength of $\lambda_\text{s}$ = 1542.4 nm. Concurrently, governed by energy conservation, its companion photon shifts to $\lambda_\text{i} = (1/\lambda_\text{p} - 1/\lambda_\text{s})^{-1}$, yielding symmetric emission peaks at 1537.6 nm for $\lambda_\text{p}$ = 770.0 nm and 1547.2 nm for $\lambda_\text{p}$ = 772.4 nm. Because these shifted companion photons are detuned by $\sim$4.8 nm away from the resonance, which far exceeds its $\sim$0.47 nm linewidth, they loses the resonant local field amplification. Since SPDC generates photons simultaneously in pairs, losing resonant enhancement for one photon directly suppresses the joint pair-generation probability, which accounts for the nearly three-order-of-magnitude reduction in the overall rate. This highlights the extreme sensitivity of the biphoton yield to the dual-resonance condition. Nevertheless, this spectral pinning mechanism guarantees exceptional wavelength stability for the primary emission channel against pump frequency fluctuations.

\section{Conclusions}

In conclusion, we have demonstrated a high-$Q$ flatband resonant 3R-MoS$_{2}$ metasurface for efficient and broad-angle quantum photon-pair generation. By implementing a lattice-constant doubling strategy to fold the Brillouin zone and introducing in-plane symmetry perturbation, we successfully mapped the dispersionless guided mode into an accessible high-$Q$ flatband resonance via the quasi-BIC mechanism at the $\Gamma$ point, preserving robust field confinement across a wide range of incident angles. The intense localized optical field enhancement within the nanoresonators, coupled with the high non-zero $\chi^{(2)}$ nonlinear tensor of 3R-MoS$_{2}$, dramatically boosts the efficiency of nonlinear optical frequency conversion via SPDC. Utilizing quantum–classical correspondence, our numerical calculations confirm that this platform achieves an significant enhancement in SPDC pair-generation rate relative to an unpatterned 3R-MoS$_{2}$ film. Crucially, the near-zero energy dispersion of the flatband resonance extends the effective SPDC emission angle up to $\pm 5^{\circ}$ with negligible spectral shift, while maintaining a high bandwidth-averaged yield under realistic pulsed pump excitation and strong spectral pinning of the resonant photon channel against pump wavelength detuning. By overcoming the angle-sensitivity and narrow acceptance limits of conventional steep-dispersion quasi-BICs, both the emission angular spectrum and pair-generation rate exhibit an order-of-magnitude improvement over previous works. Note that the principle of the precisely-engineered high-$Q$ flatband resonance can be broadly extended to enhance other nonlinear processes through suitable material and geometric scaling\cite{Sun2024, sun2025high, Zhang2025, Jiang2025, Wang2026}, and can be further augmented through integration with two-dimensional materials\cite{Liu2021, Gao2026} as well as electro-optical materials\cite{He2024, Zhang2025a} for dynamically reconfigurable quantum nanophotonics. This work establishes a versatile strategy for engineering high-brightness, angle-insensitivity quantum light sources in van der Waals nanostructures, unlocking new possibilities for compact, scan-free, wide-field quantum ghost imaging and quantum information processing.

\begin{acknowledgments}
	
This work was supported by the National Natural Science Foundation of China (Grants No. 12304420, No. 12264028, and No. 12364045), the Natural Science Foundation of Jiangxi Province (Grants No.20262BAC240253, and No. 20253BAC260002), and the Young Elite Scientists Sponsorship Program by JXAST (Grants No. 2023QT11 and 2025QT04).
	
\end{acknowledgments}

\section*{Data availability}

The data that support the findings of this article are not publicly available. The data are available from the authors upon reasonable request.


%

\end{document}